\input phyzzx.tex
\tolerance=1000
\voffset=-0.0cm
\hoffset=0.7cm
\sequentialequations
\def\rl{\rightline}

\def\t1{{\tilde 1}}

\REF{\DIM}{N. Arkani-Hamed, S. Dimopoulos and G. Dvali,  Phys. Rev. Lett.
{\bf B429} (1998) 263, hep-ph/9803315;
I.~Antoniadis, N.~Arkani-Hamed, S.~Dimopoulos and G.~Dvali,
Phys. Lett. {\bf B436}, 257 (1998), hep-ph/9804398.}
\REF{\OLD}{I. Antoniadis, Phys. Lett. {\bf B246} (1990) 377; J. Lykken,
 Phys. Rev. {\bf D54} (1996) 3693, hep-th/9603133; I. Antoniadis and K.
Benakli, Phys. Lett. {\bf B326} (1994) 69;
I. Antoniadis, K. Benakli and M. Quiros, Nucl. Phys. {\bf B331} (1994) 313.}
\REF{\EXP}{N. Arkani-Hamed, S. Dimopoulos and G. Dvali,  hep-ph/9807344.}
\REF{\CUL}{S. Cullen and M. Perelstein, hep-ph/9903422.}
\REF{\HAL}{L. Hall and D. Smith, hep-ph/9904267.}
\REF{\OTH}{K.R.~Dienes, E.~Dudas and T.~Gherghetta,
Phys. Lett. {\bf B436}, 55 (1998)
hep-ph/9803466; R.~Sundrum, hep-ph/9805471, hep-ph/9807348; G.~Shiu and
S.H.~Tye,
Phys. Rev. {\bf D58}, 106007 (1998) hep-th/9805157;
P.C.~Argyres, S.~Dimopoulos and J.~March-Russell, Phys. Lett. {\bf B441}, 96
(1998)
hep-th/9808138; N. Arkani-Hamed, S. Dimopoulos and J. March-Russell,
hep-th/9809124;
A.~Donini and S.~Rigolin, hep-ph/9901443; Z. Berezhiani and G. Dvali,
 hep-ph/9811378; Z.~Kakushadze, hep-th/9812163, hep-th/9902080; Z. Kakushadze
and S.-H.H. Tye, hep-th/9809147; N. Arkani-Hamed, S. Dimopoulos,
hep-ph/9811353;
N. Arkani-Hamed, S. Dimopoulos, G. Dvali and J. March-Russell, hep-ph/9811448;
P.~Mathews, S.~Raychaudhuri and K.~Sridhar, hep-ph/9811501; hep-ph/9812486;
A.~Pomarol and M.~Quiros, Phys. Lett. {\bf B438}, 255 (1998);
T.E.~Clark and S.T.~Love, hep-th/9901103; T. Banks, M. Dine and A. Nelson,
hep-th/9903019.}
\REF{\ACC}{G.F. Giudice, R. Rattazzi and J.D. Wells, hep-ph/9811291;
 T. Han, J.D. Lykken and R. Zhang, hep-ph/9811350;  J.L. Hewett,
hep-ph/9811356;
E.A. Mirabelli, M. Perlstein and M.E. Peskin,  hep-ph/9811337;  P. Mathews, S.
Raychaudhuri and K. Srihar,  hep-ph/9812486,  hep-ph/9811501; I. Antoniadis and
C. Bachas,
hep-th/9812093; M.L.~Graesser, hep-ph/9902310;
S. Nussinov and R.E. Shrock, hep-ph/9811323;
T.G. Rizzo, hep-ph/9901209; hep-ph/9902273; hep-ph/9903475;   K.~Agashe and
N.G.~Deshpande, hep-ph/9902263; K. Cheung, hep-ph/9904266; K. Cheung and W.
Keung,
hep-ph/9903294;
 M.~Masip and A.~Pomarol, hep-ph/9902467.}
\REF{\COS}{K. Benakli, hep-ph/9809582; K. Benakli and S. Davidson,
hep-ph/9810280; M. Maggiore and A. Riotto, hep-th/9811089; D.H. Lyth,  Phys.
Lett. {\bf B448} (1999) 191, hep-ph/9810320; N. Kaloper and A. Linde,
hep-th/9811141; G. Dvali and S.-H. H. Tye, hep-ph/9812483; K.R. Dienes, E.
Dudas, T. Ghergetta, A. Riotto,
hep-ph/9809406; N. Arkani-Hamed, S. Dimopoulos, N. Kaloper and J.
March-Russell, hep-ph/9903239.}
\REF{\ADD}{I. Antoniadis, S. Dimopoulos and G. Dvali, Nucl. Phys. {\bf B516}
(1998) 70.}
\REF{\DG}{S. Dimopoulos and G. Giudice, hep-ph/9602350.}
\REF{\ANT}{I. Antoniadis, hep-ph/9904272.}
\REF{\PRI}{J. Long, H. Chan and J. Price, Nucl. Phys. {\bf B539} (1999) 23, and
references
therein.}
\REF{\AFIV}{G. Aldazabal, A. Font, L. Ibanez and G. Violero, hep-th/9804026.}
\REF{\EDI}{E. Halyo, hep-ph/9901302.}
\REF{\SST}{G. Shiu, R. Schrock and S.-H. Tye, hep-ph/9904262.}
\REF{\IMR}{L. Ibanez,  C. Munoz and S. Rigolin, hep-ph/9812397.}
\REF{\IRU}{L. Ibanez, R. Rabadan and A. Uranga, hep-th/9808139.}

\singlespace
\rl{SU-ITP-99-20}
\rl{hep-ph/9904432}
\rl{\today}
\pagenumber=0
\normalspace
\medskip
\bigskip
\titlestyle{\bf{ Moduli Forces in String Models with Large Internal
Dimensions}}
\smallskip
\author{ Edi Halyo{\footnote*{e--mail address: halyo@dormouse.stanford.edu}}}
\smallskip
\centerline {Department of Physics}
\centerline{Stanford University}
\centerline {Stanford, CA 94305}
\smallskip
\vskip 2 cm
\titlestyle{\bf ABSTRACT}

We estimate the strength and range of forces mediated by string moduli in type
I string models with two (or more) large internal dimensions. We find that
forces mediated by twisted moduli which live on the brane world--volume can
mediate forces which are orders of magnitude stronger than gravity with a range
up to a milimeter. If they exist, these forces can be easily observed in
present experiments. On the other hand, forces mediated by
the dilaton and untwisted moduli are about a hundred times stronger than
gravity and may be observed depending on their range.

\singlespace
\vskip 0.5cm
\endpage
\normalspace

\centerline{\bf 1. Introduction}
\medskip

Recently there has been great interest in string theories at the TeV scale with
large internal dimensions[\DIM]\foot{For earlier work along this line see
[\OLD].}. In this scenario, all Standard Model (or Minimally Supersymmetric
Standard Model) degrees of freedom are confined to the $3+1$ dimensional
noncompact brane world--volume whereas gravity propagates only
in the compact bulk. The weakness of gravity with respect to the other
interactions is a result of the large size of some of the internal dimensions.
After compactification the four dimensional Newton constant is given by
$$G_N={G_{d+4} \over R^d} \eqno(1)$$
where $R$ is the large compactification radius and $G_{d+4}=M_{d+4}^{-d-2} \sim
(TeV)^{-d-2}$. Here
$d$ denotes the number of large dimensions with all other compact dimensions of
the string scale.  (In string theory $d \leq 6$ necessarily.) The most
interesting case which may also be
the easiest to experimentally confirm or rule out is the $d=2$ case.
Therefore, throughout this paper we will consider the case of two large
dimensions unless it is explicitly stated.
Surprisingly, this scenario
cannot be easily ruled out either by accelarator experiments or by
astrophysical
observations[\EXP]. The strongest bound on the higher dimensional Planck scale
scale $M_6$
comes from the supernova 1987A
giving $M_6>50~TeV$ ($M_s >4~TeV$) for two (three or more) large
dimensions[\CUL]\foot{However see [\HAL] for a model dependent higher bound of
$M_6>110~TeV$.}. In most of this paper we will assume that $M_s \sim M_6$ but
this is not necessarily the case e.g. for small string coupling. During the
last year various aspects of this scenario have been explored
including phenomenology[\OTH], collider signatures[\ACC] and cosmology[\COS].

An immediate consequence of this scenario is the change in the gravitational
force law
at a length scale which corresponds to the size of the large
dimensions[\ADD,\DG,\ANT], e.g for two large dimensions with $M_6 \sim M_s \sim
50~TeV$ this corresponds to $R \sim 0.4 \mu m$. At this scale, we expect the
gravitational force law to change from $1/r^2$ to $1/r^4$. However, this length
scale is too small for present and planned experiments to probe[\PRI]. It would
be desirable if
this scenario had an observable signature at larger length scales such as tens
of  microns which can be probed by these experiments.
In this paper, we consider forces mediated by moduli in string models with
large internal dimensions. In these models, SUSY must effectively be broken at
or below $M_s \sim TeV$ and as a result the
Compton wavelength of these moduli are rather large, e.g. in the micron range.
This is also the range of the new moduli force which can be seen as a
correction to gravity.

String models with large internal dimensions can be concretely realized in type
I orbifold compactifications with D--branes
or equivalently by type IIB orientifold compactifications[\AFIV]. In type I
string models, a priori there are three kinds of
moduli which can mediate new forces, the dilaton $S$, untwisted moduli $T_i$
and twisted  moduli $M_i$. We argue that $S$ and $T$ which are bulk fields
mediate forces with a range of microns if SUSY is broken on the brane (by some
field theoretic nonperturbative
phenomenon like gaugino condensation). However, their interactions with matter
cannot be much stronger than gravity
and therefore not easily observable by the present and planned experiments.
On the other hand, twisted moduli $M$ which live on the (orbifold) fixed planes
can
mediate forces many orders of magnitude stronger than gravity.\foot{The same
fields give rise to acceptable D--term inflation in type I string
models[\EDI].}
If the branes coincide with the fixed  planes, these moduli are confined to the
brane world--volume and they
have interactions with matter which are suppressed by $M_s$
rather than $M_P$. The integration over the large volume factor which is the
source of the $M_P$ suppression of graviton, $S$ and $T$
interactions with matter is missing in this case. Thus, with respect to gravity
their interactions with matter are enhanced by the very large factor
$(M_P/M_s)^2$. Unfortunately, due to the same enhancement
the twisted moduli also get very large i.e. $\sim M_s$ masses. As a result,
their interactions
with matter are completely negligible in the micron range. On the other hand,
these heavy twisted moduli can give rise to interesting phenomenology[\SST].

Fortunately, there are other possible configurations. For example, there may be
a fixed compact torus under the orbifold group. Then the twisted moduli also
live on this torus in addition to the brane world--volume. In this case,
the coupling of the twisted moduli to matter is suppressed by the square root
of the volume of the torus in string units. If the fixed torus is large enough
the moduli masses can be small enough to mediate a force with a macroscopic
range. In addition, the moduli
the coupling to matter can be suppressed (relative to the enhancement
$(M_P/M_s)^2$) to give forces much stronger than gravity which are not
experimentally ruled out.
In this paper, we estimate the twisted moduli masses and the strength of the
force mediated by them in generic type I string models.
We show that for realistic values of the parameters such as compactification
radii, supersymmetry (SUSY) breaking scale and $M_s$ the force mediated by $M$
can be long range and orders of magnitude
stronger than gravity.

This paper is organized as follows. In section 2 we derive the general formulas
for forces mediated by string moduli and mention how these apply to type I
string models. Section 3 is a brief review of type I string models
which realize the scenario with large internal dimensions. In section 4 we
estimate the magnitude of the modulus force and the moduli masses in a number
of different scenarios. We show that forces mediated by twisted moduli are
generically much stronger than gravity and other moduli forces.
Section 5 is a discussion of our results.

\bigskip
\centerline{\bf 2. Moduli Forces in String Theory}
\medskip

In heterotic string models with TeV scale SUSY breaking the masses of string
moduli such as
the dilaton $S$ and untwisted moduli $T_i$ is given by $m_{S,T} \sim TeV^2/M_P
\sim 10^{-12}~GeV$[\ADD].
This corresponds to a Compton wavelength of milimeters. Thus, in these models
string moduli can mediate
forces with a macroscopic range. Such a new modulus force modifies
the gravitational interaction as[\DG]
$$V(r)=-G_N{m_1m_2 \over r} \left(1+{G_{S,T}^2 } e^{-r/\lambda_{S,T}}
\right) \eqno(2)$$
where $G_{S,T}^2$ gives the relative strength of the modulus force
to gravity and
$\lambda_{S,T}$ is the range of the modulus force fixed by the Compton
wavelength
of the modulus.

The couplings of $S$ and $T$ to matter arise from the dependence of the QCD
coupling constant on the dilaton (at tree level) and the moduli (at one loop)
$${1 \over g_3^2}=k_3ReS+f(T) \eqno(3)$$
where $k_3=1$ and
$f(T)$ gives the one--loop corrections which depend on the moduli.
On dimensional grounds the nucleon mass can be written as
$$m_N \sim exp(-8\pi^2/bg_3^2) M_P \eqno(4)$$
Here $b$ is the first coefficient of the QCD $\beta$--function.
The dilaton--nucleon coupling is given by[\DG,\ADD]
$$G_S={\partial m_N \over \partial S} {\partial S \over \partial S_c}
\eqno(5)$$
with a similar expression for the modulus--matter coupling.
The second term on the right hand side takes into account the fact that moduli
generically
have nontrivial Kahler potentials and the force is mediated by the canonically
normalized
modulus field $S_c$.
Due to the large hierarchy between $M_P$ and $m_N$ the dilaton coupling to
matter is
rather large compared to gravity; e.g. $G_S^2 \sim10^3$. The modulus coupling
is much weaker than gravity, e.g. $G_T <<1$ because it arises at one loop and
is suppressed by a factor of $1/16 \pi^2$.
(However in models with Scherk--Schwarz SUSY breaking at the TeV scale the
modulus
coupling can be as strong as gravity[\ADD].)

In string models with large internal dimensions the situation is slightly
different.
First, the string scale (or rather $M_{d+4}$) is not large but around the TeV
scale. Taking into account
the bounds from supernova 1987A we assume $M_6 \sim M_s> 50~TeV$. If SUSY is
broken
only on the brane this gives a Compton wavelength
of $\lambda_{S,T} <0.4 \mu m$ for two large dimensions. Thus, the range of the
modulus force is much shorter than a milimeter. Second, not every modulus
necessarily couples to observable matter. In the heterotic string models
mentioned above this was the case due to the fact that both $S$ and $T$
appeared in the expression for $g_3$. We will see that in type I string models
either $S$ or $T$ but not both couple to matter.
Moreover, in these models the couplings of $S$ and $T$ to matter are more than
an order of magnitude weaker than the corresponding forces in heterotic models
since there are only four to five orders of magnitude between $m_N$ and $M_s$.
We conclude that in this scenario forces mediated by untwisted moduli such as
$S$ and $T$ are short range and not strong enough and therefore
probably cannot be observed in present or planned experiments. (Note however
that for small string coupling or for more than two large dimensions a lower
string scale is possible. Then, the range of the force would be longer and it
may be observable.)

Finally, there can be twisted moduli which cannot propagate in the bulk as $S$
and $T$ but live on the (orbifold) fixed planes. If the branes are at the fixed
point interactions of twisted moduli with matter are suppressed only by $M_s$
rather than $M_P$.
As a result, their couplings to matter are many orders of magnitude stronger
than gravity. Unfortunately, in this case moduli masses are unsuppressed and
$\sim M_s$. Therefore these
forces are completely negligible at the micron range. However, if there is a
fixed torus of the orbifold group
the moduli coupling to matter and their masses are suppressed by the square
root of the volume of the torus. For a large enough torus this may lead to
strong moduli interactions with a long range.
In these cases the twisted moduli may have
long range interactions which are orders of magnitude stronger than gravity.
In type I string models there are twisted moduli $M$ with the above properties.
Their
VEVs are related to the blowup of the orbifold  singularities which smoothes
out
the singular compactification geometry. In section 4 we will estimate
the force mediated by these twisted moduli and their masses for a number of
different scenarios.

\bigskip
\centerline{\bf 3. Type I String Theory or Orientifolds of Type IIB Strings}
\medskip

Four dimensional type I string compactifications with $N=1$ supersymmetry can
be obtained by orientifolds of type IIBstrings[\AFIV]. We start with a type IIB
string theory in $D=10$ and mode it out by the world--sheet parity
transformation $\Omega$. This gives a type I string theory in $D=10$ with gauge
group
$SO(32)$. The gauge group arises from the 32 D9 branes required for tadpole
cancellation.
This type I string theory is further compactified on an orbifold of $T^6=T^2
\times T^2 \times T^2$
(with radii $R_{1,2,3}$)
i.e. on $T^6/\Gamma$ where $\Gamma$ is a discrete group such as $Z_n$ or $Z_n
\times Z_m$
resulting in a $D=4$ theory with $N=1$ supersymmetry and chiral matter
content. The above construction has only D9 branes but by considering more
elaborate orientifolds one can obtain models with two types of branes in the
theory.
For example, if $O_i$ denotes reflection of  $T^2_i$ then one can mode $T^6$ by
$(-1)^{F_L} \Omega O_i$, $\Omega O_i O_j$ or $(-1)^{F_L} \Omega O_i O_j O_k$.
Here $F_L$ is the left
handed world--sheet fermion number. The above modings result in D7, D5 and D3
branes respectively.
Due to the requirement for $N=1$ SUSY only two sets of branes can appear
simultaneously; either D9 and D5 branes
or D3 and D7 branes[\AFIV].
We denote D5 branes as $5_i$ if the branes wrap around
$T^2_i$ and the D7 branes as $7_i$ if they wrap around $T^2_j \times T^2_k$.
The number of each kind of brane is fixed again by tadpole cancellation. On the
two different kinds of branes ($Dp$ and $Dp^{\prime}$) there are gauge
multiplets from strings with both ends on the same kind of brane (i.e. $pp$ or
$p^{\prime} p^{\prime}$ strings), giving two gauge groups, $G_p$ and
$G_{p^\prime}$. Different polarizations of the same open strings also give
matter multiplets in the adjoint representation of the gauge group.
In additon, there are matter fields which arise from strings with ends on
different kinds of branes (i.e. $pp^{\prime}$ strings) in the bifundamental
representation.
Newton's constant is given by
$$G_N={1 \over M_P^2}= {g_I^2 \over {8M_I^8 R_1^2 R_2^2 R_3^2}} \eqno(6)$$
Here $g_I$ is the type I string coupling constant and $M_I=
\alpha_{str}^{-1/2}$ is the type I string scale.

We now consider the two possibilities, i.e. models with D3--D7 branes and
D9--D5 branes separately[\AFIV].

a) Models with D3 and D7 branes: These are obtained by moding out type IIB
string theory by the group generated by $(-1)^{F_L} \Omega O_i$ and $(-1)^{F_L}
\Omega O_i O_j O_k$.
There are two gauge groups $G_3$ and $G_7$ living on the D3 and D7 branes.
The matter content of these models arises from strings stretched between
different branes, i.e. 33, 37, 73 and 77 strings (Here and below we omit the
indices $i,j$ for notational simplicity.). We denote these fields by
$M^{33},M^{37},M^{73},M^{77}$.
$M^{33}$ and $M^{77}$ are in the adjoint representation of the respective gauge
groups, $G_3$
and $G_7$, whereas $M^{37}$ and $M^{73}$ are in the bifundamental
representation. In realistic models there are Wilson lines which break the
gauge group and project out part of the matter. The tree level superpotential
generically contains the terms[\IMR]
$$W=g_3(M^{33} M^{73} M^{37})+g_7(M^{77} M^{37} M^{73}) \eqno(7)$$
The gauge couplings are given by ($ \alpha_i=g_i^2/4 \pi$)
$$\alpha_3={g_I \over 2}  \qquad \alpha_{7_i}={g_I \over {2M_I^4 R_j^2 R_k^2}}
\eqno(8)$$

There are three kinds of moduli; the dilaton $S$ and the untwisted moduli $T_i$
are defined as
$$S= {2  \over g_I} +i \theta \eqno(9)$$
and
$$T_i={2 R_j^2 R_k^2 M_I^4\over g_I}+i \eta_i \eqno(10)$$
where $\theta$ and $\eta_i$ are untwisted Ramond--Ramond fields.
The Kahler potential for $S$ and $T_i$ is given by
$$K(S, \bar S, T_i, \bar T_i)=-log(S+ \bar S)-\Sigma_i log(T_i+ \bar T_i)
\eqno(11)$$
The third kind of moduli are the twisted moduli denoted by $M_i=\phi_i+i
\psi_i$[\AFIV]. $\phi_i$ and $\psi_i$ are twisted NS--NS and R--R fields
respectively.
The VEV of $\phi_i$ parametrize the blowup of the orbifold singularities
or the smooothing of the compact space. The Kahler potential for $M_i$ is not
known but we will assume that it is canonical.
In the presence of $M$ the connection between the gauge couplings and the
moduli are given by[\AFIV]
$${1 \over \alpha_3}=Re S+s_3 M \eqno(12)$$
and
$${1 \over \alpha_{7_i}}=Re T_i+s_{7_i} M \eqno(13)$$
Here the $s_{3,7_i}$ are constants of $O(1-10)$. In these models, there is
fixed torus $T^2_j \times T^2_k$ of the of the orbifold group $O_i$. Therefore,
twisted moduli corresponding to this twist
live in eight dimensions along the D7 branes. The orbifold group $O_iO_jO_k$
has only fixed points and therefore
twisted moduli corresponding to these live in four dimensions parallel to the
D3 branes.

b) Models with D5 and D9 branes: These are obtained by moding out type IIB
string theory by
the group generated by $\Omega$ and $\Omega O_i O_j$. The generic gauge and
matter
content and the superpotential are very similar to the previous case with the
substitution $3 \to 9$ and $5_i \to 7_i$. The reason for this is the fact that
the two kinds of models are related to each other by T duality along all the
compact dimensions. In this case the gauge couplings are
$$\alpha_9={g_I \over {M_I^6 R_1^2 R_2^2 R_3^2}} \qquad \alpha_{5_i}={g_I \over
{M_I^2 R_i^2}} \eqno(14)$$
Now, the dilaton and untwisted moduli are defined as
$$S={2 M_I^6 R_1^2 R_2^2 R_3^2 \over g_I}+i\theta \eqno(15)$$
and
$$T_i={2 M_I^4 R_j^2 R_k^2 \over g_I}+i \eta_i \eqno(16)$$
The Kahler potential for $S$ and $T_i$ remains the same. The dependence of the
$G_9$ and $G_5$ gauge couplings on the moduli is also given by eqs. (12) and
(13) with the substitution $3 \to 9$ and $5_i \to 7_i$.
Once again there are twisted moduli $M_i$ which are related to the blowup of
the singular compact space. Twisted moduli which correspond to the twist
$\Omega$ live in ten dimensions,
i.e. they can move in the whole space. The ones arising from $O_iO_j$ live in
six dimensions
since there is only a fixed $T^2_k$ under this twist.

\bigskip
\centerline{\bf  4. Moduli Forces in String Models with Large Dimensions}
\medskip

In this section we estimate the strength of the moduli forces in the scenario
with two large dimensions within the framework of
type I string theory. We assume that $M_s \sim M_6 \sim 50~TeV$
and the SUSY breaking scale is $\Lambda \sim 10~TeV$. We take the SUSY breaking
scale to be less  than the string scale in order not to generate a new
hierarchy. However it is easy to show that our results hold also for the case
$\Lambda \sim M_s$.
Using eq. (1) we find that in this case the two large compact dimensions are of
the size $R \sim 4 \times 10^{8} ~GeV^{-1} \sim 0.4 \mu m$. So there are two
$T^2$s of size $M_s^{-2}$ and one of size $R^2$.

First consider the models
with D3 and D7 branes.  We assume that we live on the D3 branes therefore we
need $g_3 \leq 1$. Thus from eq. (8) we have $g_I \leq 1/10$ (or $S \sim 20$)
i.e. we are in perturbative regime of type I string theory.
We also assume that there is one other set of D7 branes wrapping the $T^2
\times
T^2$ of string scale. (Otherwise there are D7 branes wrapeed around the large
dimensions and therefore light Kaluza--Klein states on the branes which are not
desirable.) From eq. (10) we find that one of the three untwisted moduli is
around $T \sim 2/g_I$ whereas the other two are very large, $T_i \sim 4 \times
10^{26}$.

In models with D9 and D5 branes the situation is similar. Once again we have D9
branes and
one set of D5 branes wrapping a $T^2$ of the string size. Now from eq. (15) we
have
$S \sim 4 \times 10^{26}$ giving a very small $g_9$. Thus, we assume that our
world lives on the three noncompact dimensions of the D5 brane
world--volume. In this case there two possibilities. If the two large
dimensions belong to the same $T^2$ then there are two untwisted moduli with
$T_i \sim 2/g_I$ and one with $T \sim 4 \times
10^{26}$. On the other hand, if the two large dimensions belong to two
different $T^2$s then
there is one modulus with $T \sim 2/g_I$ and two large moduli of size $T_i \sim
2 \times 10^{13}$.

In these models there are two possibilities for SUSY breaking. SUSY can be
broken either on the brane or in the bulk. (Of course it can also be broken
both in the bulk and on the brane but this case is obtained from the two
above.)
If SUSY is broken on the brane by a mechanism like gaugino condensation
at a scale $\Lambda$ then this breaking is mediated to the observable sector
either by gauge interactions or by nonrenormalizble interactions suppressed by
powers of $M_s$. Then $m_{S,T} \sim \Lambda^3/M_s M_P$ since interactions of
the bulk modes with the brane are suppressed by $M_P$. This is a small mass and
therefore $S$ and $T$ have the potential to mediate macroscopic range forces.
On the other hand, if SUSY is broken in the bulk by an F--term of $O(M_sM_P)$
then  $S$ and $T$ get very large i.e. $\sim M_S$ masses.
Therefore their interactions with matter are very short range and completely
negligible at the micron range.

We can now estimate the strength of the forces mediated by the untwisted moduli
when SUSY is broken on the brane only.  The moduli masses are $m_{S,T} \sim
\Lambda^3/M_s M_P \sim  2 \times10^{-11}~GeV$ giving a range of $\lambda_{S,T}
\sim 50 \mu m$.
In models with D3 and D7 branes $T$ does not couple to observable matter since
it does not appear in the expression for $g_3$. $S$ on the other hand
does couple to matter. Due to the nontrivial Kahler potential for $S$ given by
eq. (11) the
canonical dilaton is given by $S_c \sim logS$ (at least for the real part of
$S$ which we are interested in). Using $M_s \sim 50~TeV$ and eq. (4) for $m_n$
and eq. (5) we get
$G_s \sim 10$. Thus the dilaton force in this case is about two orders of
magnitude stronger than
gravity with a range of $50 \mu m$. In models with D9 and D5 branes our world
lives on the D5 branes and therefore
$S$ does not couple to matter. The only untwisted modulus $T$ that appears in
$g_3$
is the one with a small VEV, i.e. $T \sim 2/g_I$. The canonical modulus field
is given by
$T_c \sim log T$ and the analysis is identical to that of $S$ above resulting
in a force stronger than gravity
mediated by the untwisted modulus. If the range of the modulus force is $\sim
50 \mu m$ then it becomes of the order of gravity around $200 \mu m$ which is
around what present experiments can detect. We conclude that for this choice of
parameters forces mediated by $S$ and $T$ are about two orders of magnitude
stronger than gravity and may be observed. Of course, if we assume
$\Lambda \sim M_s$ the range of the force becomes much smaller e.g. $\sim 0.4
\mu m$ and then it cannot be observed by the present experiments.
(However, note that we assume $M_s \sim M_6$ here. If the string coupling is
small
or there are more than two large dimensions so that $M_s \sim 3-10~TeV$ then
this
force with $G^2_{S,T} \sim 100$ has a range around $10-100 \mu m$ which can
easily be detected.)
Above the strength of the $S$ and $T$ moduli forces is normalized by the Newton
constant
since these fields live in the bulk. As a result, their couplings to matter
which live on the brane are suppressed by the large volume factor, i.e. $M_P$.

We now turn to the twisted moduli $M$ which are present in these models.
The coupling of $M$ to matter arises from the dependence of $g_3$ on $M$ given
by eq. (12).
Then the nucleon mass becomes
$$m_N \sim exp(-8\pi^2/bg_3^2) M_s \sim exp[-8\pi^2 Re(S+s_i
M)/b]M_s\eqno(17)$$
In some models $s_i=b \sim 10$, in any case these constants are of the same
magnitude[\IRU]. Therefore, for simplicity in our estimates we will take them
to be equal. Note that the coefficient
$s_i$ is may be a large number (in which case the VEV of $M$ has to be smaller
than $M_s$).
As we will see this is the origin of the strong coupling of $M$ to matter. If
$s_i \sim1$ coupling $M$ to matter would be as strong as those of $S$ or $T$
only.
The twisted moduli $M$ live on the fixed planes and cannot propagate in the
bulk.
If the D--branes coincide with the fixed planes the twisted moduli live on the
branes. As a result, their interactions
with matter are suppressed by $M_s$ and not $M_P$ giving an enhancement of
$(M_P/M_s)^2 \sim 4 \times 10^{26}$ with respect to gravity. Using eq. (17) for
$m_N$ we find that $G_M \sim (2s/b) \times 10^{15}$ which gives a force thirty
orders of magnitude stronger than gravity at its peak. The mass of $M$ is
independent of where SUSY is broken. If SUSY is broken on the brane then $m_M
\sim M_s$ and the moduli force has a very short range. On the other hand, if
SUSY is broken in the bulk with
$F \sim M_s M_P$ (in order to give large masses to superparticles) then again
$m_M \sim M_s$.
Therefore, when the branes overlap with the fixed planes, even though the
modulus force is very strong at its peak, it is negligible at the micron range
due to its extremely short range.

We found that the modulus mass is generically $\sim M_s$ leading to a very
short range for the force mediated by $M$. Our arguments above are valid for
the orbifolds with twists that have
only fixed points but no fixed tori in the compact directions. If there are
such fixed tori then the untwisted moduli can move on them in addition to the
noncompact brane world--volume.
For example, consider an orbifold twist with a fixed $T^2$ such as $(-1)^F
\Omega O_iO_j$ we
considered in the previous section for type I string models. Then the
corresponding twisted modulus lives in six dimensions including the compact
torus $T^2_k$ which is invariant under this twist. As a result, its four
dimensional effective interactions with matter living on the brane
are suppressed by the square root of the volume of the torus in string units.
We find that the interaction with matter is given by
$$G_M \sim \left({8 \pi^2 g_I^2 s\over b}\right) \left({M_P \over M_s}\right)
\left({1 \over {R^2M_s^2}}\right)^{1/2} \eqno(18)$$
 and the modulus mass is
$$m_M \sim g_I \left({\Lambda^3 \over M_s^2}\right)  \left({1 \over
{R^2M_s^2}}\right)^{1/2} \eqno(19)$$
Now, if the fixed torus has large dimensions there is a large suppression in
the modulus mass
which can give a long range for the modulus force. Since the force is thirty
orders of magnitude stronger than gravity before this suppression we expect it
to still be much stronger than gravity
with the volume suppression. It is easy to see that if the fixed torus has two
large dimensions
$$G_M \sim \left({8 \pi^2 g_I^2 s\over b}\right) \eqno(20)$$
Thus in this case we find $G_M \sim 80$ giving a force $6 \times 10^3$ times
stronger than gravity (for the realistic choice of parameters $s \sim b$, $g_I
\sim 1$). We stress that the force
mediated by $M$ is stronger than those mediated by $S$ or $T$ due to large
value of $s_i \sim b
\sim 10$. Also it is easy to see that the volume suppression of the interaction
is exactly the factor
that converts $1/M_s^2$ to $1/M_P^2$ so that the force is enhanced only by the
factor in eq. (20).

The range of the force is given by the Compton wavelength of the modulus
$\lambda_M \sim M_s M_P /\Lambda^3$ which is always larger than the size of the
large dimension $R \sim M_P/M_s^2$ for $\Lambda<M_s$. In the above case
$\lambda_M \sim 50 \mu m$. The force has range of $\sim 100R$ which is about
the range of the force mediated by $S$ or $T$. It becomes of gravitational
strength
around $450 \mu m$ which is easily observable by the present experiments. This
is an effect much more dominant than the change in the gravitational
force or the force mediated by untwisted moduli.

Above we assumed that the SUSY breaking scale $\Lambda$ is smaller than $M_s$
in order not to generate a new hierarchy problem in light of the large string
scale. However one can show
that even if $\Lambda \sim M_s$ the modulus force is much stronger than
gravity.
The strength of the moduli force at its peak is given by $G_M^2 \sim (8\pi^2 s
g_I^2/b)^2$
and is independent of the SUSY breaking scale or the range of the moduli force.
When
$\Lambda \sim M_s$, however the range of the force $\lambda_M$ is equal to the
size of the large dimension $R$.
For  many reasonable values of the parameters $G_M^2$ above is much larger than
unity giving a force a few orders of magnitude stronger than gravity.

We also assumed that $M_s \sim M_6$ which is not the case for small string
coupling. The force mediated by twisted string moduli can be much stronger than
gravity even if  $M_s<M_6$.
For example, take $M_s \sim 10~TeV$ and $\Lambda \sim 5 ~TeV$
for $M_6 \sim 50~TeV$. The size of the large dimensions
is $R \sim 10 \mu m$ whereas the range of the modulus force is $\lambda_M \sim
80 \mu m$ for two large dimensions. The strength of the force at this distance
is $G_M^2 \sim 6 \times10^3$. This
force becomes of the gravitational strength at $\sim 0.8 mm$.
Thus we find that even if $M_s<M_6$ there can be moduli forces which are much
stronger than gravity which are easily observable.

Most of the above analysis was done for two large dimensions for which there is
a model independent bound on the six dimensional Planck scale $M_6>50~TeV$. For
more than  two large dimensions this bound relaxes significantly, i.e. for
three large dimensions $M_7>4~TeV$ and for four and more large dimensions
$M_d>1~TeV$[\CUL]. It is easy to show that the strength of the force and its
range are independent of the number of large dimensions. Assuming that the
twisted moduli live in $4+d$ dimensions the range of the force
(for  $d$ large dimensions) is given by
$$\lambda_M \sim \left(M_s^2 \over \Lambda^3 \right) \left( 1\over
{R^dM_s^d}\right)^{-1/2}
\eqno(21)$$
where $R$ is given by eq. (1) to be $R^d \sim M_P^2/M_s^{d+2}$. We see that the
range is
$\lambda_M \sim M_sM_P/ \Lambda^3$ independent of the number of large
dimensions.
In particular for $\Lambda \sim M_s$ we find $\lambda_M \sim M_P/ M_s^2$.
On the other hand, the strength of the moduli force in $d$ dimensions is given
by
$$G_M \sim \left({8 \pi^2 g_I^2 s\over b}\right) \left({M_P \over M_s}\right)
\left( 1\over {R^dM_s^d}\right)^{1/2}
\eqno(22)$$
We find that for all $d$, $G_M \sim (8 \pi^2 g_I^2 s/ b)$ giving a force $\sim
6 \times 10^3$ times
stronger than gravity.
We see that independent of the number of large dimensions if $\Lambda \sim M_s$
then the range of the force $\lambda_M \sim R$. However, the the moduli force
is about three orders
of magnitude stronger than gravity at this scale and therefore a much larger
effect.

For example, consider the case of three large dimensions with $M_s \sim M_7
\sim \Lambda \sim 4~TeV$[\CUL]. Then the large radii are of size $60 \mu m$ and
the range of the moduli force is also $\lambda_M \sim 60 \mu m$ (and even
longer if we take $\Lambda<M_s$). Now $G_M^2 \sim (8 \pi^2)^2
\sim 6500$ and it becomes of order unity at $\sim 0.6 mm$. Thus the effect of
the moduli force
is much larger than the change in gravity and easily observable in present
experiments.
With more than three large dimensions the bound on $M_d \sim M_s$ is relatively
small and all effects occur around the milimeter scale. Since forces
of gravitational strength are experimentally excluded above a milimeter it is
difficult to have moduli forces stronger than gravity at this scale. However,
it is clear that if for $d \geq 3$ the string scale can be pushed up by
collider or astrophysical constraints it will be easy to have moduli forces
stronger than gravity.

Of course for more than two large dimensions, the twisted moduli do not have to
live on all the large dimensions. For example there can be three large
dimensions but the same orbifold twist considered above would still have only a
fixed $T^2$. Then $M$ live in six dimensions as before;
however now the size of the large dimension $R$ is different. In this case we
have
$$G_M \sim \left({8 \pi^2 g_I^2 s\over b}\right) \left({M_P \over
M_s}\right)^{1/3}
\eqno(23)$$
Thus we get $G_M^2 \sim 10^{10}$. The range of the force is
$$\lambda_M \sim \left(M_s^2 \over \Lambda^3 \right) \left(M_s \over M_P
\right)^{2/3}
\eqno(24)$$
This gives $\lambda_M \sim 0.1 \mu m$ which is rather small. We find that this
possibility leads to a very strong force with a very short range which is
negligible compared to the forces mediated by the untwisted moduli. However,
the three large dimensions do not have to be isotropic. For example, if the
area of the fixed $T^2$ is $R_1 R_2 \sim  10^{20} ~GeV^{-2}$
(with the third large dimension of size $R_3 \sim 80~GeV$) we find that the
range of the force
is $\lambda_M \sim 10^{10}~GeV^{-1} \sim 10 \mu m$ and the strength of the
force is given by
$G_M^2 \sim 3 \times 10^5$. We see that the anisotropic large dimensions have
the potential to give rise to very strong forces with a long range.

We implicitly considered untwisted moduli fields $M$ with canonical Kahler
potentials.
However, the Kahler potential for $M$ is probably not canonical but of the form
$$K(M_i, \bar M_i)={M_i \bar M_i \over {(T_i+ \bar T_i)^{n_i}}} \eqno(25)$$
Here $n_i$ are positive integers and $T_i$ are the untwisted moduli.
In this case, we have observable forces only if the moduli $T_i$ in the Kahler
potential have small i.e. $O(1)$ values that is they correspond to radii of
string size. Above we saw that in all models there are such untwisted moduli.
Otherwise, there is a very large field renormalization due to
the above Kahler potential $M$ is renormalized by a factor of $(2 \times
10^{13})^{n_i}$. This makes the already very strong modulus force many orders
of magnitude stronger; however it also renormalizes the modulus mass by the
same factor. As a result, the force becomes extremely short range and therefore
completely negligible at the micron range.

\bigskip
\centerline{\bf 5. Conclusions}

In this paper we have shown that type I string models with large internal
dimensions may have light moduli which can mediate forces orders of magnitude
stronger than gravity with a macroscopic range. We found that the untwisted
moduli $S$ and $T$ can mediate forces about $100$ times
gravity with a range of  microns. On the other hand, the twisted moduli $M$
which live on the branes
can mediate forces orders of magnitude stronger than gravity with a range that
is longer than the size of the large dimensions.
As a result, these effects are much easier to probe by the present and planned
experiments than
the change in the gravitational force. This is especially the case for two
large dimensions with
a string scale of $50~TeV$ which puts the size of the large dimensions out of
experimental reach.

The smallness of  the moduli masses is the reason for the long range of the
force. The mass is small due to the suppression of the moduli interactions with
matter by the square root of the volume of a large torus on which the fixed
moduli can move. The same suppression factor also
reduces the huge enhancement in the moduli interactions with matter. In fact if
the twisted moduli move on all the large dimensions this factor takes $1/M_s$
to $1/M_P$ giving gravitational strength interactions up to the enhancement
coming from the moduli coupling to
the nucleons. We saw that if the string parameter $s_i \sim b$ the force is
orders of magnitude stronger than gravity. We find that there can
be moduli forces with a range of tens to hundreds of microns and strength of
four orders of magnitude stronger than gravity for a wide range of parameters
of the string model. If the orbifold
group of the string model does not have fixed tori or these tori are string
size then the range
of the moduli force is extremely small and it cannot be detected.

The above results are quite robust. They hold for realistic values of the
string scale, SUSY breaking scale and other parameters of the string model.
In particular, they are valid for any number of large internal dimensions as we
showed above. The strength of the moduli interactions with matter is
independent of the SUSY breaking scale and the number of dimensions (but
depends on the bound on the string scale and on $s_i$). The range of the force
given by the Compton wavelength of the moduli is independent of the number of
large dimensions but is strongly dependent on the SUSY breaking scale and the
string scale.

Of course, the presence of these moduli forces is not a model independent
prediction. There may be string models with discrete torsion in which most or
all of the twisted moduli are projected out. In addition, even if there are
such twisted moduli in a given model they may obtain masses of
$\sim M_s$ in which case such forces are completely negligible. However, it is
interesting to see that there are generic scenarios with such forces with
experimentally observable effects.

\bigskip
\centerline{\bf Acknowledgements}
I would like to thank Jaume Gomis and Nemanja Kaloper for very useful
discussions.

\vfill

\refout

\end
\bye